\documentclass[12pt]{article}
\usepackage{amssymb}

\def\hybrid{\topmargin 0pt      \oddsidemargin 0pt
        \headheight 0pt \headsep 0pt
        \voffset=-0.5cm
        \hoffset=-0.25in
        \textwidth 6.75in
        \textheight 9.5in       
        \marginparwidth 0.0in
        \parskip 5pt plus 1pt   \jot = 1.5ex}
\catcode`\@=11
\def\marginnote#1{}

\newcount\hour
\newcount\minute
\newtoks\amorpm
\hour=\time\divide\hour by60 \minute=\time{\multiply\hour by60
\global\advance\minute by-\hour}
\edef\standardtime{{\ifnum\hour<12 \global\amorpm={am}%
        \else\global\amorpm={pm}\advance\hour by-12 \fi
        \ifnum\hour=0 \hour=12 \fi
        \number\hour:\ifnum\minute<10 0\fi\number\minute\the\amorpm}}
\edef\militarytime{\number\hour:\ifnum\minute<10 0\fi\number\minute}

\def\draftlabel#1{{\@bsphack\if@filesw {\let\thepage\relax
   \xdef\@gtempa{\write\@auxout{\string
      \newlabel{#1}{{\@currentlabel}{\thepage}}}}}\@gtempa
   \if@nobreak \ifvmode\nobreak\fi\fi\fi\@esphack}
        \gdef\@eqnlabel{#1}}
\def\@eqnlabel{}
\def\@vacuum{}
\def\draftmarginnote#1{\marginpar{\raggedright\scriptsize\tt#1}}
\def\draftlabel#1{{\@bsphack\if@filesw {\let\thepage\relax
   \xdef\@gtempa{\write\@auxout{\string
      \newlabel{#1}{{\@currentlabel}{\thepage}}}}}\@gtempa
   \if@nobreak \ifvmode\nobreak\fi\fi\fi\@esphack}
        \gdef\@eqnlabel{#1}}
\def\@eqnlabel{}
\def\@vacuum{}
\def\draftmarginnote#1{\marginpar{\raggedright\scriptsize\tt#1}}

\def\draft{\oddsidemargin -.5truein
        \def\@oddfoot{\sl preliminary draft \hfil
        \rm\thepage\hfil\sl\today\quad\militarytime}
        \let\@evenfoot\@oddfoot \overfullrule 3pt
        \let\label=\draftlabel
        \let\marginnote=\draftmarginnote
   \def\@eqnnum{(\theequation)\rlap{\kern\marginparsep\tt\@eqnlabel}%
\global\let\@eqnlabel\@vacuum}  }

\def\numberbysection{\@addtoreset{equation}{section}
        \def\theequation{\thesection.\arabic{equation}}}

\def\underline#1{\relax\ifmmode\@@underline#1\else
        $\@@underline{\hbox{#1}}$\relax\fi}

\def\titlepage{\@restonecolfalse\if@twocolumn\@restonecoltrue\onecolumn
     \else \newpage \fi \thispagestyle{empty}\c@page\z@
        \def\thefootnote{\fnsymbol{footnote}} }

\def\endtitlepage{\if@restonecol\twocolumn \else  \fi
        \def\thefootnote{\arabic{footnote}}
        \setcounter{footnote}{0}}  

\numberbysection \hybrid

\newcommand{\Rf}{{\mathbf R}}
\newcommand{\Tf}{{\mathbf T}}

\newcommand{\ps}{\mathsf{p}}

\newcommand{\g}{{\mathbf g}}

\newcommand{\e}{{\mathbf e}}

\newcommand{\mC}{\mathbb C}
\newcommand{\mZ}{\mathbb Z}

\newtheorem{ex}{Example}[section]

\newtheorem{predl}{Proposition}[section]

\newtheorem{rem}{Remark}[section]

\def\beq{\begin{equation}}
\def\eeq{\end{equation}}
\def\p{\partial}

\def\res{\mathop{\hbox{Res}}\limits}

\begin{document}

\setcounter{page}{1}

\

\vspace{-18mm}

\begin{flushright}
 ITEP-TH-32/18\\
\end{flushright}
\vspace{0mm}

\begin{center}
\vspace{5mm}
{\LARGE{Supersymmetric extension of}}
 \\ \vspace{4mm}
{\LARGE{qKZ--Ruijsenaars correspondence}}

 \vspace{18mm}

 {\large  {A. Grekov}\,\footnote{Moscow
Institute of Physics and Technology, Inststitutskii per.  9,
Dolgoprudny, Moscow region, 141700, Russia; ITEP,
B.Cheremushkinskaya 25, Moscow 117218, Russia; Steklov Mathematical
Institute of Russian Academy of Sciences, Gubkina str. 8, 119991,
Moscow, Russia; Skolkovo Institute of Science and Technology, 143026
Moscow, Russian Federation; e-mail: grekovandrew@mail.ru}
 \quad\quad\quad
 {A. Zabrodin}\,\footnote{National Research University
Higher School of Economics, Russian Federation; Institute of
Biochemical Physics of Russian Academy of Sciences, Kosygina str. 4,
119334, Moscow, Russian Federation; Skolkovo Institute of Science
and Technology, 143026 Moscow, Russian Federation; e-mail:
zabrodin@itep.ru}
 \quad\quad\quad
 {A. Zotov}\,\footnote{Steklov Mathematical Institute of Russian
Academy of Sciences, Gubkina str. 8, 119991, Moscow, Russia; ITEP,
B.Cheremushkinskaya 25, Moscow 117218, Russia; National Research
University Higher School of Economics, Russian Federation; Moscow
Institute of Physics and Technology, Inststitutskii per.  9,
Dolgoprudny, Moscow region, 141700, Russia; e-mail: zotov@mi-ras.ru}
 }
\end{center}

\vspace{10mm}

 \begin{abstract}
We describe the correspondence of the Matsuo-Cherednik type between
the quantum $n$-body Ruijsenaars-Schneider model and the quantum
Knizhnik-Zamolodchikov equations related to supergroup $GL(N|M)$.
The spectrum of the Ruijsenaars-Schneider Hamiltonians is shown to
be independent of the $\mZ_2$-grading for a fixed value of $N+M$, so
that $N+M+1$ different qKZ systems of equations lead to the same
$n$-body quantum problem. The obtained results can be viewed as a
quantization of the previously described quantum-classical
correspondence between the classical $n$-body Ruijsenaars-Schneider
model and the supersymmetric $GL(N|M)$ quantum spin chains on $n$
sites.
 \end{abstract}

\newpage




\section{Introduction}
\setcounter{equation}{0}

{\em The KZ-Calogero and qKZ-Ruijsenaars correspondences} are the
Matsuo-Cherednik type constructions
\cite{Matsuo-Cherednik,FV,ZZ1,ZZ2} for solutions of the
Calogero-Moser-Sutherland \cite{Calogero} and
Ruijsenaars-Schneider \cite{RS} quantum problems by means of
solutions of the Knizhnik-Zamolodchikov (KZ) \cite{KZ} and quantum
Knizhnik-Zamolodchikov (qKZ) equations \cite{qKZ} respectively.
Consider, for example, the qKZ equations\footnote{The quantum
$R$-matrices entering (\ref{qkz1a}) are assumed to be unitary: ${\bf
R}_{ij}(x){\bf R}_{ji}(-x)=\mbox{id}$.} related to the Lie group
$GL(K)$:
 \beq\label{qkz1}
e^{\eta \hbar \p_{x_i}} \Bigl | \Phi \Bigr >={\bf K}_i^{(\hbar )}
\Bigl | \Phi \Bigr >, \qquad i=1, \ldots , n,
 \eeq
 \beq\label{qkz1a}
{\bf K}_i^{(\hbar )}= {\bf R}_{i\, i\! -\! 1}(x_i \! -\! x_{i-1}\!
+\! \eta \hbar )\ldots {\bf R}_{i 1}(x_i \! -\! x_{1}\! +\! \eta
\hbar ) {\bf g}^{(i)} {\bf R}_{i n}(x_i \! -\! x_{n})\ldots {\bf
R}_{i\, i\! +\! 1}(x_i \! -\! x_{i+1} )\,,
 \eeq
where ${\bf g}=\mbox{diag}(g_1, \ldots , g_K)$ is a diagonal $K\!
\times \! K$ (twist) matrix, and ${\bf g}^{(i)}$ acts by ${\bf g}$
multiplication in the $i$-th tensor component of the Hilbert space
${\cal V}=(\mC^K)^{\otimes n}$. The quantum $R$-matrices ${\bf
R}_{ij}$ are in the fundamental representation of $GL(K)$.
They act in the $i$-th and $j$-th tensor components of ${\cal V}$
and satisfy the quantum Yang-Baxter equation, which guarantees
compatibility of equations (\ref{qkz1}).
 The twist
matrix ${\bf g}$ is the symmetry of ${\bf R}_{ij}$: ${\bf
g}^{(i)}{\bf g}^{(j)}{\bf R}_{ij}={\bf R}_{ij}{\bf g}^{(i)}{\bf
g}^{(j)}$. In the rational case we deal with the Yang's $R$-matrix
\cite{Yang}:
  \beq\label{r1}
{\bf R}_{ij}(x)=\frac{x{\bf I} +\eta {\bf P}_{ij}}{x+\eta},
 \eeq
where ${\bf I}$ is identity operator in ${\rm End}(\cal V)$, and
${\bf P}_{ij}$ is the permutation operator, which interchanges the
$i$-th and $j$-th tensor components in $\cal V$.  The
operators\footnote{The set $\{e_{ab}\,|\,a,b=1...K\}$ is the
standard basis in ${\rm Mat}(K,\mC)$:
$(e_{ab})_{ij}=\delta_{ia}\delta_{jb}$.}
 \beq\label{Ma}
{\bf M}_a=\sum_{l=1}^{n}e_{aa}^{(l)}
 \eeq
commute with ${\bf K}_i^{(\hbar )}$ and provide the weight
decomposition of the Hilbert space $\cal V$ into the direct sum
 \beq\label{weight1}
 \displaystyle{
{\cal V}= V^{\otimes n} \,\, = \bigoplus_{M_1, \ldots , M_K} \!\!
\!\! {\cal V}(\{M_a \})
 }
 \eeq
 of eigenspaces of operators
${\bf M}_a$ with the eigenvalues $M_a \in \mZ_{\geq\, 0}$, $a=1,
\ldots , K$: $M_1 +\ldots +M_K=n$. Using the standard basis
$\{e_a\}$ in $\mC^K$ introduce the basis vectors in ${\cal V}(\{M_a
\})$ as the vectors
 \beq\label{q1}
 \displaystyle{
 \Bigl |J\Bigr > =e_{j_1}\otimes
e_{j_2}\otimes \ldots \otimes e_{j_n},
 }
 \eeq
 where the number of indices
$j_k$ such that $j_k =a$ is equal to $M_a$ for all $a=1, \ldots ,
K$. The dual vectors $\Bigl <J\Bigr |$ are defined in so that
$\Bigl <J\Bigr |J'\Bigr
>=\delta_{J, J'}$.

Then the statement of the qKZ-Ruijsenaars correspondence is as
follows \cite{ZZ2}. For any solution of the qKZ equations
(\ref{qkz1}) $\displaystyle{\Bigl |\Phi \Bigr >=\sum_J \Phi_J \Bigl
|J\Bigr >}$ from the weight subspace ${\cal V}(\{M_a \})$ the
function
 \beq\label{qkz2}
\Psi =\sum_J \Phi_J\,,\quad  \Phi_J=\Phi_J(x_1,...,x_n)
 \eeq
or
 \beq\label{qkz20}
\Psi =\Bigl <\Omega \Bigr | \Phi \Bigr >\,,\quad\quad
 \Bigl < \Omega \Bigr |=\sum\limits_{J:\,\bigl |J \bigr >\in\, {\cal V}(\{M_a
\}) } \Bigl < J \Bigr |
 \eeq
 with the property
  \beq\label{qkz23}
 \Bigl <\Omega \Bigr |{\bf P}_{ij}=\Bigl <\Omega \Bigr |
 \eeq
is an eigenfunction of the Macdonald difference operator:
 \beq\label{qkz3}
\sum_{i=1}^n  \prod_{j\neq i}^n \frac{x_i\! -\! x_j \! +\! \eta}{x_i
-x_j}\, \Psi (x_1, \ldots , x_i\! +\! \eta \hbar , \ldots , x_n)
=E\Psi (x_1, \ldots , x_n)\,,\quad\quad E=\sum_{a=1}^K M_a g_a\,.
 \eeq
The eigenvalues of the higher rational Macdonald-Ruijsenaars
Hamiltonians
 \beq\label{q2}
  \displaystyle{
  \hat {\cal H}_d=\sum\limits_{I\subset\{1, \ldots , n\}, |I|=d}\Bigl(
  \prod\limits_{s\in I,r\not\in I}\frac{x_s-x_r+\eta}{x_s-x_r}\Bigr)
  \prod\limits_{i\in I}e^{\eta\hbar\p_{x_i}}
  }
 \eeq
 are given by the elementary symmetric polynomial of
$n$ variables
 $e_d(\underbrace{g_1, \ldots , g_1}_{M_1},\, \ldots \,
\underbrace{g_N, \ldots , g_K}_{M_K})$.

{\em QC-duality.} Using the asymptotics of  solutions to the (q)KZ
equations \cite{TVa} it was also argued in \cite{ZZ1,ZZ2} that the
qKZ-Ruijsenaars correspondence can be viewed as a quantization of the
quantum-classical duality \cite{AKLTZ13,GZZ,BLZZ} (see also
\cite{MTV,GK}), which relates the generalized inhomogeneous quantum
spin chains and the classical Ruijsenaars-Schneider model. Consider
the classical $K$-body Ruijsenaars-Schneider model, where the
positions of particles $\{x_i\}$ are identified with the
inhomogeneity parameters of the spin chain which is described by its
transfer matrix
 \beq\label{q3}
 {\bf T}(x)=\mbox{tr}_0\Bigl (\widetilde {\bf R}_{0n}(x-x_n)\ldots
\widetilde {\bf R}_{02}(x-x_2) \widetilde {\bf R}_{01}(x-x_1)({\bf
g}\otimes {\bf I})\Bigr )
 \eeq
with the $R$-matrix
 \beq\label{q4}
 \widetilde {\bf R}(x)= \frac{x+\eta}{x}\, {\bf R}(x) ={\bf
I}+\frac{\eta}{x}\, {\bf P}.
 \eeq
The quantum spin chain Hamiltonians are defined as follows:
 \beq\label{q5}
{\bf H}_i= \res\limits_{x=x_i}{\bf T}(x)=\widetilde {\bf R}_{i\, i\!
- \! 1}(x_i \! -\! x_{i-1})\ldots \widetilde {\bf R}_{i 1}(x_i \!
-\! x_{1}) {\bf g}^{(i)} \widetilde {\bf R}_{i n}(x_i \! -\!
x_{n})\ldots \widetilde {\bf R}_{i\, i\! +\! 1}(x_i \! -\! x_{i+1}
).
 \eeq
 Therefore,
  \beq\label{q6}
 {\bf H}_i={\bf K}_i^{(0)}\prod_{j\neq i}^n
\frac{x_i-x_j+\eta}{x_i-x_j}\,,\quad {\bf K}_i^{(0)}={\bf
K}_i^{(\hbar)}\left.\right|_{\hbar=0}\,.
 \eeq
Identify also the generalized velocities $\{{\dot x}_i\}$ with the
eigenvalues of (\ref{q5}). Then the action variables
$\{I_i|\,i=1,...,K\}$ of the classical model (eigenvalues of the Lax matrix)
are given by the values
of $g_1,...,g_K$ with multiplicities $M_1,...,M_K$:
  \beq\label{q7}
\{I_i|\,i=1,...,K\}=\Big\{\underbrace{g_1, \ldots , g_1}_{M_1},\,
\ldots \, \underbrace{g_N, \ldots , g_K}_{M_K}\Big\}\,.
 \eeq
See details in \cite{GZZ}, where this statement was proved using the
algebraic Bethe ansatz technique.

{\em QC-correspondence.}  On the other hand, the quantum-classical
duality possesses a generalization to the so-called
quantum-classical correspondence \cite{TZZ}, where the classical
Ruijsenaars-Schneider model is related not to a single spin chain
but to the set of $K+1$ supersymmetric spin chains \cite{Kulish}
associated with supergroups
  \beq\label{q8}
GL(K|\,0)\,,\ GL(K-1|\,1)\,,\quad...\quad\,,
GL(1|\,K-1)\,,\ GL(0|\,K)\,.
 \eeq
More precisely, it was shown in \cite{TZZ} that the previous
statement (\ref{q7}) is valid for all supersymmetric chains with
supergroups (\ref{q8}).

{\em The aim of this paper} is to quantize the (supersymmetric)
quantum-classical correspondence, that is to establish
supersymmetric version of the qKZ-Ruijsenaars correspondence for the
qKZ equations related to the supergroups $GL(N|M)$. We
construct generalizations of the vector $\Bigl < \Omega \Bigr |$
(\ref{qkz20}) and show that the quantum $K$-body
Ruijsenaars-Schneider model follows from all $K+1$ qKZ systems of
equations related to the supergroups $GL(N|M)$ with $N+M=K$
(\ref{q8}). The skew-symmetric vectors $\Bigl < \Omega_- \Bigr |$
with the property $\Bigl < \Omega_- \Bigr |{\bf P}_{ij}=-\Bigl <
\Omega_- \Bigr |$  (instead of symmetric vector (\ref{qkz23})) are
described as well. They lead to the Ruijsenaars-Schneider model with
different sign of the coupling constant $\eta$ and $\hbar$.

 The paper is organized as follows. For simplicity we start with the
 rational KZ-Calogero correspondence. Then we proceed to the rational and
 trigonometric qKZ-Ruijsenaars relations. Most of notations are
 borrowed from \cite{ZZ1,ZZ2,TZZ}. We briefly describe the notations
 and definitions related to the graded Lie algebras (groups) in the
 Appendix.

\section{SUSY KZ-Calogero correspondence}
\setcounter{equation}{0}
The rational Knizhnik-Zamolodchikov (KZ) equations \cite{KZ} have the form
 \beq\label{kz1}
\hbar \p_{x_i}\Bigl | \Phi \Bigr >=\left ({\bf g}^{(i)}+ \kappa
\sum_{j\neq i}^n \frac{{\bf P}_{ij}}{x_i-x_j}\right ) \Bigl | \Phi
\Bigr >\,,
 \eeq
 where $\Bigl | \Phi \Bigr >=\Bigl | \Phi \Bigr >(x_1, \ldots , x_n)$
belongs to the tensor product ${\cal V}=V\otimes V\otimes \ldots
\otimes V =V^{\otimes n}$ of the vector spaces $V=\mathbb{C}^{N|M}$,
${\bf P}_{ij}$ is the (graded) permutation operator (\ref{a5}) of
the $i$-th and $j$-th tensor components, ${\bf g}=\mbox{diag}(g_1,
\ldots , g_{N+M})$ is a diagonal $(N+M)\! \times \! (N+M)$ matrix
and ${\bf g}^{(i)}$ is the operator in ${\cal V}$ acting as ${\bf
g}$ on the $i$-th component (and identically on the rest of the
components). The operators
 \begin{equation}
 {\bf H}_i = {\bf g}^{(i)}+
\kappa \sum_{j\neq i}^n \frac{{\bf P}_{ij}}{x_i-x_j}
 \end{equation}
 form the commutative set of Gaudin Hamiltonians \cite{Gaudin}.
Similarly to non-supersymmetric case they also commute with the
 operators:
 \begin{equation}
{\bf M}_a=\sum_{l=1}^{n}\e_{aa}^{(l)}\,,
 \end{equation}
 where $\e_{ab}$ are basis elements of ${\rm End}(\mC^{N|M})$
 (\ref{a1})-(\ref{a333}).
 In what follows we restrict ourselves to the subspace ${\cal V}(\{M_a \})$ corresponding to a component of decomposition
 (\ref{weight1}) with the fixed set of eigenvalues $M_a$ for the operators ${\bf M}_a$.
We fix a basis in ${\cal V}(\{M_a \})$: $$ \Bigl |J\Bigr >
=e_{a_1}\otimes e_{a_2}\otimes \ldots \otimes e_{a_n} = \Bigl
|a_1...a_n\Bigr >, $$ where $e_a$ are basis vectors
in $V$ and the number of indices $a_k$ such that
$a_k =a$ is equal to $M_a$ for all $a=1, \ldots , N+M$.
 A general solution to (\ref{kz1}) can be written as
 \begin{equation}
\Bigl | \Phi \Bigr > = \sum_J \Phi_J  \Bigl |J\Bigr >\,,
 \end{equation}
 where the coefficients $\Phi_J $ are functions of all parameters
 entering (\ref{kz1}).

 To proceed further we need to find a (co)vector
 \begin{equation}
\Bigl < \Omega \Bigr |=\sum_J \Bigl < J \Bigr | \Omega_J
 \end{equation}
similar to (\ref{qkz20}) with the property
 \begin{equation}\label{OMG}
\Bigl < \Omega \Bigr |{\bf P}_{ij} = \Bigl < \Omega \Bigr |\,,
 \end{equation}
 where in contrast to (\ref{qkz23}) the permutation operator ${\bf
 P}_{ij}$ acts in the graded space (it has the form (\ref{a5})).
 Having such a vector and taking into account the identities (\ref{a9})
and (\ref{a10}),
we can repeat all the calculations from \cite{ZZ1} without any
changes. They lead to the eigenvalue
 equation for the second Calogero-Moser Hamiltonian:
 \beq\label{kz3}
\left ( \hbar^2 \sum_{i=1}^n \p_{x_i}^2-\sum_{i\neq j}^n
\frac{\kappa (\kappa -\hbar )}{(x_i-x_j)^2}\right )\Psi = E\Psi\,,
 \eeq
where
 \beq\label{q91}
\Psi = \Bigl < \Omega \Bigr |\Phi \Bigr > = \sum_J \Omega_J \Phi_J
 \eeq
and
 \begin{equation}\label{q92}
E = \sum_{a=1}^{N+M} M_a g_a^2\,.
 \end{equation}
Let us construct the vector $ \Bigl < \Omega \Bigr|$. Due to
(\ref{a60}) the basis vector $ \Bigl < J \Bigr|$ entering $ \Bigl <
\Omega \Bigr|$ can not contain two identical fermions (vectors $e_{a}$
 with $\ps (a) = 1$). Otherwise we get a contradiction with (\ref{OMG}).
Keeping this in mind choose a vector $ \Bigl|J \Bigr>$ with $a_1
\leq a_2 \leq ... \leq a_n$ from ${\cal V}(\{M_a \})$, and fix the
coefficient $\Omega _ {a_1 \leq a_2 \leq ... \leq a_n} = 1$ for this
set. Next, generate the rest of vectors $ \Bigl|J \Bigr>$ by the
rule that the permutation of two nearby indices multiplies the
coefficient by the standard parity factor:
 \begin{equation}\label{q93}
\Omega _ {a_1 \; a_2  ... a_{m+1}\; a_m  ...  a_n} =
(-1)^{\ps(a_m) \ps(a_{m+1})}\Omega _ {a_1 \; a_2 ... a_m \; a_{m+1} ...  a_n}
 \end{equation}
By repeating this procedure and summing up all the resultant vectors
$ \Bigl|J \Bigr>$ (in the orbit
 of the action of permutation operators with the corresponding coefficients $\Omega_J$)
  we get the final answer for $ \Bigl| \Omega \Bigr>$. Here are some
  examples.
 \begin{ex}
Let $N+M = 2,\, n=3,\, M_1=2, M_2 = 1$, $\ps (1)=0, \ps (2)=1$. Then
  \beq\label{q351}
 \begin{array}{l}
  \displaystyle{
\Bigl| \Omega \Bigl> = \Bigl| 112 \Bigl> + \Bigl| 121 \Bigl> +
\Bigl| 211 \Bigl>\,.
 }
  \end{array}
 \eeq
 \end{ex}
 \begin{ex} Let $N+M = 3,\, n=3,\, M_1= M_2= M_3 = 1$. Then
  \beq\label{q350}
 \begin{array}{l}
  \displaystyle{
\Bigl| \Omega \Bigl> = \Bigl| 123 \Bigl> + (-1)^{\ps(1)\ps(2)}
\Bigl| 213 \Bigl> + (-1)^{\ps(2)\ps(3)} \Bigl| 132 \Bigl> +
  }
  \\ \ \\
\displaystyle{
 +
(-1)^{\ps(1)\ps(3)+\ps(2)\ps(3)} \Bigl| 312 \Bigl> +
(-1)^{\ps(1)\ps(2)+\ps(1)\ps(3)} \Bigl| 231 \Bigl> +
  }
  \\ \ \\
\displaystyle{ +(-1)^{\ps(1)\ps(2)+\ps(2)\ps(3)+\ps(1)\ps(3)} \Bigl|
321 \Bigl>\,.
   }
  \end{array}
 \eeq
 \end{ex}
\begin{ex} Let $N+M = 3,\, n=4,\, M_1=2, M_2= M_3 = 1, \ps(1)=0, \ps(2)=\ps(3) =
1$. Then
  \beq\label{q352}
 \begin{array}{l}
  \displaystyle{
\Bigl| \Omega \Bigl> = \Bigl| 1123 \Bigl> + \Bigl| 1213 \Bigl> +
\Bigl| 2113 \Bigl> + \Bigl| 1231 \Bigl> +
   }
  \\ \ \\
\displaystyle{ + \Bigl| 2311 \Bigl> + \Bigl| 2131 \Bigl> + \Bigl|
2113 \Bigl> - (2 \leftrightarrow 3) }\,.
  \end{array}
 \eeq
 \end{ex}
Note that in the case when $\ps (a)=0$ for all $a$ we return back to the
non-supersymmetric case: $\Omega_J =1$ for all $J$. On the other
hand, when $\ps (a)=1$ for all $a$ we get completely antisymmetric tensor
$\Omega_{a_1 ... a_n} = \epsilon_{a_1 ... a_n}$. Thus different
choices of $\mathfrak{B}$ (\ref{a0}) provide different
eigenfunctions (\ref{q91}). At the same time the eigenvalues are the
same (\ref{q92}), so that we get a degeneracy of the spectrum for
the Hamiltonian (\ref{kz3}).

It is also worth noting that in order to change the sign
 of $\kappa$ in the Hamiltonian (\ref{kz3}) we need to construct vector $\Bigl| \Omega_- \Bigl>$, which
  is antisymmetric under the action of permutations:
 \begin{equation}\label{OMG2}
\Bigl < \Omega_- \Bigr |{\bf P}_{ij} = -\Bigl < \Omega_- \Bigr |\,,
 \end{equation}
where the sign is opposite to the one in (\ref{OMG}). Such a vector
can not contain two identical bosons because the permutation of them
contradicts  assumption (\ref{OMG2}).
In other situations it can be constructed. The example is given below.
\begin{ex}
Let $N+M = 3,\, n=3,\, M_1= M_2= M_3 = 1$ as in (\ref{q350}) and
$\ps(1)=\ps(2)=\ps(3)=1$. Then
 \beq\label{q410}
\Bigl| \Omega_- \Bigl> = \Bigl| 123 \Bigl> + \Bigl| 213 \Bigl>
 +  \Bigl| 132 \Bigl> +  \Bigl| 312 \Bigl> + \Bigl| 231 \Bigl>  + \Bigl| 321
 \Bigl>\,.
 \eeq
 \end{ex}

\section{SUSY qKZ-Ruijsenaars correspondence: rational case }
\setcounter{equation}{0}
 In this section we generalize the correspondence between KZ
 equations and Calogero-Moser systems to
the case of SUSY qKZ equations and the
Ruijsenaars-Schneider systems. The qKZ equations have the form
  \beq\label{qkz111}
e^{\eta \hbar \p_{x_i}} \Bigl | \Phi \Bigr >={\bf K}_i^{(\hbar )}
\Bigl | \Phi \Bigr >, \qquad i=1, \ldots , n\,,
  \eeq
 where the operators in the r.h.s
  \beq
{\bf K}_i^{(\hbar )}=
{\bf R}_{i\, i\! -\! 1}(x_i \! -\! x_{i-1}\! +\! \eta \hbar )\ldots
{\bf R}_{i 1}(x_i \! -\! x_{1}\! +\! \eta \hbar ) {\bf g}^{(i)}
{\bf R}_{i n}(x_i \! -\! x_{n})\ldots
{\bf R}_{i\, i\! +\! 1}(x_i \! -\! x_{i+1} )
  \eeq
are constructed by means of the quantum $R$-matrix ${\bf R}$, which
is a (unitary) solution of the graded Yang-Baxter equation. We start
with the rational one
  \beq\label{r111}
{\bf R}_{ij}(x)=\frac{x{\bf I} +\eta {\bf P}_{ij}}{x+\eta},
  \eeq
where ${\bf P}_{ij}$ is the graded permutation operator (\ref{a5}).
Similarly to the non-supersymmetric case introduce the rescaled
$R$-matrix:
   \beq\label{r2}
\widetilde {\bf R}(x)=
\frac{x+\eta}{x}\, {\bf R}(x) ={\bf I}+\frac{\eta}{x}\, {\bf P}.
  \eeq
The transfer matrix of the corresponding supersymmetric spin chain
  \beq
\Tf (x)=\mathrm{str}_{0} \Bigl (\widetilde{\Rf} _{0n}(x-x_{n})
\ldots \widetilde{\Rf} _{02}(x-x_{2}) \widetilde{\Rf} _{01}(x-x_{1})
\, (\g \otimes \mathbf{I})\Bigr )
 \label{Top0}
  \eeq
 provides non-local Hamiltonians as its residues:
   \beq\label{r4}
{\bf T}(x)=\mbox{str} \,{\bf g}\cdot {\bf I}+\sum_{j=1}^n\frac{\eta
{\bf H}_j}{x-x_j}.
  \eeq
Explicitly,
 \beq\label{r6}
{\bf H}_i=
\widetilde {\bf R}_{i\, i\! - \! 1}(x_i \! -\! x_{i-1})\ldots
\widetilde {\bf R}_{i 1}(x_i \! -\! x_{1}) {\bf g}^{(i)}
\widetilde {\bf R}_{i n}(x_i \! -\! x_{n})\ldots
\widetilde {\bf R}_{i\, i\! +\! 1}(x_i \! -\! x_{i+1} ).
  \eeq
Alternatively,
   \beq\label{r5}
{\bf H}_i={\bf K}_i^{(0)}\prod_{j\neq i}^n
\frac{x_i-x_j+\eta}{x_i-x_j}\,.
  \eeq
 From comparison of expansions of the transfer matrix as $x\to
 \infty$ in the forms (\ref{Top0}) and (\ref{r4})
 \begin{equation}
\mbox{str}\, {\bf g} \cdot {\bf I}+\frac{\eta}{x}\sum_{i=1}^n
\mbox{str}_0 \Bigl ( {\bf P}_{0i}{\bf g}^{(0)}\Bigr )+\ldots = \mbox{str}\, {\bf g} \cdot {\bf I}+\frac{\eta}{x}\sum_{i=1}^n
{\bf H}_i+\ldots
 \end{equation}
we obtain:
  \beq\label{r7}
\sum_{i=1}^{n}{\bf H}_i= \sum_{i=1}^n {\bf g}^{(i)}=
\sum_{a=1}^{N+M}g_a {\bf M}_a,
  \eeq
where the property (\ref{a10}) was used. To obtain the
correspondence we project the qKZ-equations on the vector $\Bigl |
\Omega \Bigr >$ (\ref{OMG}), constructed in the previous section:
 \begin{equation}
e^{\eta \hbar \p_{x_i}}\Bigl <\Omega \Bigr |\Phi \Bigr >=
e^{\eta \hbar \p_{x_i}}\Psi =
\Bigl <\Omega \Bigr |   {\bf K}_{i}^{(\hbar )}\Bigl | \Phi \Bigr >=
\Bigl <\Omega \Bigr |   {\bf K}_{i}^{(0 )}\Bigl | \Phi \Bigr >.
 \end{equation}
and repeat all calculations from \cite{ZZ2}. This yields:
 $$
\sum_{i=1}^n \left ( \prod_{j\neq i}^n \frac{x_i-x_j
+\eta}{x_i-x_j}\right ) e^{\eta \hbar \p_{x_i}}\Psi = \sum_{i=1}^n
\prod_{j\neq i}^n \frac{x_i-x_j +\eta}{x_i-x_j}\, \Bigl <\Omega
\Bigr |   {\bf K}_{i}^{(0 )}\Bigl | \Phi \Bigr >
 $$
 $$
=\sum_{i=1}^n \Bigl <\Omega \Bigr |   {\bf H}_{i}\Bigl | \Phi \Bigr
> = \sum_{i=1}^n \Bigl <\Omega \Bigr |   {\bf g}^{(i)}\Bigl |
\Phi \Bigr >=\sum_{a=1}^{N+M} g_a \Bigl <\Omega \Bigr | {\bf
M}_a\Bigl | \Phi \Bigr >=\left (\sum_{a=1}^{N+M} g_a M_a\right
)\Psi\,,
 $$
where
  \begin{equation}\label{r12}
\Psi = \Bigl<\Omega \Bigr |\Phi \Bigr >
 \end{equation}
is the eigenfunction and
 \begin{equation}
E = \sum_{a=1}^{N+M} g_a M_a
 \end{equation}
is the eigenvalue.
\begin{rem}
To obtain the Macdonald-Ruijsenaars Hamiltonian with the
 opposite sign of the coupling constant $\eta$ and $\hbar$ one should start with the $R$-matrix
 \begin{equation}\label{r113}
{\bf R}_{ij}(x)=\frac{x{\bf I} +\eta {\bf P}_{ij}}{x-\eta}
 \end{equation}
 in (\ref{qkz111}) instead of (\ref{r111}).
The $R$-matrix (\ref{r113}) is still unitary and acts identically on
the antisymmetric vector $\Bigl | \Omega_- \Bigr >$ (\ref{OMG2})
which is to be used instead of $\Bigl | \Omega \Bigr >$.
\end{rem}
\subsubsection*{Higher Hamiltonians} Following the
construction in the non-supersymmetric case,
it can be shown that the wave function $\Psi=\Bigl<\Omega \Bigl |
\Phi \Bigr>$ satisfies the equations
  \beq\label{a3}
  \displaystyle{
\prod\limits_{s=1}^d e^{\eta\hbar\frac{\p\ }{\p x_{i_s}}}\Psi=\Bigl
<\Omega\Bigr |
 {\bf K}_{i_1}^{(0 )}\ldots{\bf K}_{i_d}^{(0  )} \Bigr | \Phi \Bigr
 >\quad \hbox{for}\quad i_k\neq i_m\,.
  }
  \eeq
The proof of this statement is the same as in \cite{ZZ2}.
 One more point needed for the correspondence is the determinant identity
  \beq\label{a7a}
\det_{1\leq i,j \leq n}\left (z\delta_{ij}-\frac{\eta {\bf H}_i}{x_j-x_i+\eta}\right )=
\prod_{a=1}^{N}(z-g_a)^{{\bf M}_a}.
 \eeq
It was proven for the supersymmetric case in \cite{TZZ}. Therefore,
the correspondence works in the supersymmetric case as well. Namely,
given a solution $|\Phi \Bigr >$ of the qKZ equations the wave
function of the rational Ruijsenaars-Schneider quantum problem is
given by (\ref{r12}). The eigenvalues are the same symmetric
polynomials as in the non-supersymmetric case (\ref{q2}).

\section{SUSY qKZ-Ruijsenaars correspondence, trigonometric case}
\setcounter{equation}{0}
 The trigonometric (hyperbolic) solution to the graded
Yang-Baxter equation has the following form \cite{BR}:
 \beq\label{q411}
   \begin{array}{c}
   \displaystyle{
{\bf R}_{12}(x)=\frac{1}{2 \sinh (x \! + \! \eta )} \! \sum_{a=1}^{N+M} \left(e^{x+\eta}
q^{-2\ps(a)} - e^{-x-\eta} q^{2\ps (a)}\right) e_{aa}\otimes e_{aa}
+\frac{\sinh x}{\sinh (x + \eta )} \sum_{a\neq b}^{N+M}
e_{aa}\otimes e_{bb}
 }
 \\ \ \\
  \displaystyle{
+ \frac{\sinh \eta}{\sinh (x + \eta )}\sum_{a<b}^{N+M} \Bigl ( e^x
(-1)^{\ps (b)} e_{ab}\otimes e_{ba}+e^{-x} (-1)^{\ps (a)}
e_{ba}\otimes e_{ab}\Bigr )\,,
  }
  \end{array}
 \eeq
where $q = e^\eta$. It can be  rewritten as follows:
 \beq\label{q412}
{\bf R}_{12}(x)\! = {\bf P}_{12} + \frac{\sinh x}{\sinh (x\! +\!
\eta )}\Bigl ({\bf I}-{\bf P}^q_{12} \Bigr ) + {\bf G}_{12}^+\,,
 \end{equation}
where ${\bf P}_{12}$ is the graded permutation operator (\ref{a5}),
 ${\bf P}^q_{12}$ -- its $q$-deformation (the quantum permutation operator)
 \begin{equation}\label{q413}
{\bf P}^q_{12} =\sum_{a=1}^{N+M} (-1)^{\ps(a)} e_{aa}\otimes e_{aa}
+q\sum_{a>b}^{N+M} (-1)^{\ps(b)}e_{ab}\otimes e_{ba}
+q^{-1}\sum_{a<b}^{N+M} (-1)^{\ps(b)} e_{ab}\otimes e_{ba}
 \end{equation}
and
 \beq\label{q414}
   \begin{array}{c}
      \displaystyle{
{\bf G}_{12}^+ = \sum_{a=1}^{N+M} \Bigl( \frac{\sinh(x+\eta-2\eta
\ps(a))}{\sinh(x+\eta)}
 -(-1)^{\ps(a)} +\frac{\sinh(x)}{\sinh(x+\eta)}((-1)^{\ps(a)}-1)\Bigr )e_{aa}\otimes e_{aa}
 }
 \\ \ \\
   \displaystyle{
=2\sum_{a \in \mathfrak{F} } \frac{ (\cosh \eta - 1)
\sinh x}{\sinh(x+\eta)} \,
  e_{aa}\otimes e_{aa}
 }
  \end{array}
 \eeq
or
 \beq\label{q415}
   \begin{array}{c}
      \displaystyle{
{\bf G}_{12}^+ =  \sum_{a=1}^{N+M} {\bf G}_a^+ \, e_{aa}\otimes
e_{aa}\,,\quad\quad {\bf G}_a^+=\frac{ (1-(-1)^{\ps(a)})(\cosh \eta
- 1) \sinh x}{\sinh(x+\eta)}\,.
 }
  \end{array}
 \eeq
The $R$-matrix entering the transfer matrix differs from
(\ref{q411}) by a scalar factor:
  \beq\label{tr4}
{\widetilde {\bf R}}_{12}(x)= \frac{\sinh (x+\eta )}{\sinh x}\, {\bf
R}_{12}(x)\,,
  \eeq
and the transfer matrix itself is defined similarly to (\ref{Top0}).
 The Hamiltonians are introduced through the expansion
  \begin{equation}
{\bf T}(x)={\bf C}+\sinh \eta \sum_{k=1}^n {\bf H}_k \coth
(x-x_k)\,.
 \end{equation}
They are related to the operators in the r.h.s of the qKZ-equations
by the same formulae as in non-supersymmetric case:
  \beq\label{tr6a}
{\bf H}_i={\bf K}_i^{(0)}\prod_{j\neq i}^n \frac{\sinh(x_i-x_j+\eta
)}{\sinh (x_i-x_j)}\,.
  \eeq

\subsubsection*{Construction of $q$-symmetric vectors}

Our strategy is as follows. Following the non-supersymmetric
construction \cite{ZZ2}, we now need to find a vector $\Bigl
<\Omega_q \Bigr |$ with the property
  \beq\label{tr9}
\Bigl <\Omega_q \Bigr |\, {\bf R}_{i\, i\! -\! 1}(x)=\Bigl <\Omega_q
\Bigr |\, {\bf P}_{i\, i\! -\! 1}, \qquad i=2, \ldots , n\,.
  \eeq
Let us show that this vector has the form:
 \begin{equation}\label{q417}
\Bigl < \Omega _q\Bigr |=\sum_J q^{\ell (J)}\Omega_J\Bigl < J \Bigr
|\,,
 \end{equation}
where $\Omega_J$ are the same as in the rational case (\ref{kz3}),
(\ref{q93}), while $\ell (J)$
 is defined to
 be the minimal number of elementary permutations
 required to get the
multi-index $J=(j_1, j_2, \ldots , j_n)$ starting from the
``minimal''
 one.
 The ``minimal'' order implies that the $j_k$'s are ordered as $1\leq j_1\leq j_2\leq
\ldots \leq j_n\leq N$ (see \cite{ZZ2}).
The proof is straightforward. First, by the construction we see that
 \begin{equation}
\Bigl <\Omega_q \Bigr |{\bf P}^q_{i,i\! -\! 1}=\Bigl <\Omega_q \Bigr
|\,.
 \end{equation}
In contrast to the non-supersymmetric case we have additional terms
${\bf G}_{i,i-1}^+$ in $R$-matrices (\ref{q412}). However, they do
not provide any effect when acting on $\Bigl <\Omega_q \Bigr |$:
 \begin{equation}
\Bigl <\Omega_q \Bigr |{\bf G}_{i,i-1}^+  = 0\,.
 \end{equation}
It happens because of the tensor structure (\ref{q414}). Indeed,
 \begin{equation}
{\bf G}_{i,i-1}^+ \Bigl| J \Bigr > = {\bf G}_{a_i}^+ \delta_{a_i,
a_{i-1}} \Bigl| J \Bigr >\,,
 \end{equation}
so that only the same basis vectors $e_{a_i}$ entering $\Bigl| J
\Bigr
>$ may contribute. But we have already assumed that our vector
$\Bigl <\Omega_q \Bigr |$ does not contain two identical fermions, and for bosons
${\bf G}_{a}^+ = 0$. Finally, using (\ref{q412}) we arrive at
(\ref{tr9}).

\begin{ex} Let $N+M = 3,\, n=3,\, M_1= M_2= M_3 = 1$. Then
  \beq\label{q439}
   \begin{array}{l}
   \displaystyle{
\Bigl| \Omega_q \Bigl> = \Bigl| 123 \Bigl> + q \,
(-1)^{\ps(1)\ps(2)} \Bigl| 213 \Bigl> + q\,(-1)^{\ps(2)\ps(3)}
\Bigl| 132 \Bigl> +
  }
   \\ \ \\
  \displaystyle{
 + q^2(-1)^{\ps(1)\ps(3)+\ps(2)\ps(3)} \Bigl| 312 \Bigl> + q^2\,(-1)^{\ps(1)\ps(2)+\ps(1)\ps(3)} \Bigl| 231 \Bigl> +
  }
   \\ \ \\
  \displaystyle{
  +q^3(-1)^{\ps(1)\ps(2)+\ps(2)\ps(3)+\ps(1)\ps(3)} \Bigl| 321
  \Bigl>\,.
   }
  \end{array}
 \eeq
 \end{ex}

\subsubsection*{Calculation of the eigenvalue}

Coming back to the proof of the correspondence  we need the identity
 \begin{equation}
\Bigl <\Omega_q \Bigr |{\bf K}_{i}^{(\hbar )}=
 \Bigl <\Omega_q \Bigr |   {\bf K}_{i}^{(0 )} = \Bigl <\Omega_q\Bigr |\, {\bf P}_{i\, i\! -\! 1}\ldots {\bf
 P}_{i1}\,,
 \end{equation}
 which follows from
 $
{\bf P}_{i\, i\! -\! 1} {\bf P}^q_{i\, i\! -\! 2}={\bf P}^q_{i\! -\!
1 \, i\! -\! 2}{\bf P}_{i\, i\! -\! 1}
 $
and an analogue of the identity $$\displaystyle{ {\bf T}(\pm \infty
)={\bf C}\pm \sinh \eta \sum_k {\bf H}_k =\sum_{a=1}^{N} g_ae^{\pm
\eta {\bf M}_a}}
$$
 for the supersymmetric case. It is as follows.
 \begin{predl}
 \beq\label{q419}
   \begin{array}{l}
   \displaystyle{
{\bf T}( \infty ) = \sum_{a \in \mathfrak{B}} g_a e^{ \eta {\bf
M}_a} - \sum_{a \in \mathfrak{F}} g_ae^{ -\eta {\bf M}_a},
  }
 \\ \ \\
  \displaystyle{
  {\bf T}(
-\infty ) = \sum_{a \in \mathfrak{B}} g_a e^{ -\eta {\bf M}_a} -
\sum_{a \in \mathfrak{F}} g_ae^{ \eta {\bf M}_a}.
   }
  \end{array}
 \eeq
 \end{predl}
\noindent\underline{\em{Proof:}} $\quad$  We will prove the first
equality. The proof of the second one
is similar. Let us first find the asymptotics
of the $R$-matrix:
 \beq\label{q420}
   \begin{array}{c}
   \displaystyle{
  \widetilde{{\bf R}}(\infty) = {\bf I} + (q-q^{-1})\sum_{a<b}^{N+M} (-1)^{\ps(b)}
  e_{ab}\otimes e_{ba}
   + (q-1)\sum_{a=1}^{N+M} (-1)^{\ps(a)} e_{aa}\otimes e_{aa}
 }
 \\ \ \\
  \displaystyle{
   +  \sum_{a=1}^{N+M}\Bigl (q^{1-2\ps(a)}- (-1)^{\ps(a)}q+ ((-1)^{\ps(a)}-1) \Bigr) e_{aa}\otimes
   e_{aa}\,.
  }
  \end{array}
 \eeq
  This expression can be rewritten in the following form:
     \beq
 \widetilde{{\bf R}}(\infty) = {\bf I} + (q-q^{-1})
 \sum_{a<b}^{N+M} (-1)^{\ps(b)} e_{ab}\otimes e_{ba} +
 \sum_{a=1}^{N+M}\Bigl (q^{1-2\ps(a)}-1 \Bigr ) e_{aa}\otimes
 e_{aa}\,.
   \eeq
  The off-diagonal part does not contribute to the trace in
  (\ref{Top0}). Therefore,
    \beq\label{q425}
   \begin{array}{l}
   \displaystyle{
 {\bf T}( \infty ) = \sum_{a=1}^{N+M} (-1)^{\ps(a)} g_a \prod_{j=1}^{n} \Bigl (1 + (q^{1-2\ps(a)}-1)e_{aa}^{(j)}  \Bigr) =
 }
 \\ \ \\
  \displaystyle{
  =
 \sum_{a=1}^{N+M} (-1)^{\ps(a)} g_a \prod_{j=1}^{n}
 \Bigl (1 + \sum_{N_j =1}^{\infty} \frac{\eta^{N_j} (1-2\ps(a))^{N_j}}{N_j !}
 e_{aa}^{(j)}  \Bigr) =
 }
 \\ \ \\
  \displaystyle{
 =
 \sum_{a=1}^{N+M} (-1)^{\ps(a)} g_a \prod_{j=1}^{n} \Bigl
 (\sum_{N_j =0}^{\infty} \frac{\eta^{N_j} (1-2\ps(a))^{N_j}}{N_j !} (e_{aa}^{(j)})^{N_j}
 \Bigr)
 }
  \end{array}
 \eeq
and, finally,
 \beq\label{q426}
   \begin{array}{c}
  \displaystyle{
 {\bf T}( \infty ) =
 \sum_{a=1}^{N+M} (-1)^{\ps(a)} g_a \prod_{j=1}^{n} \Bigl
 ( e^{\eta (1-2\ps(a))e_{aa}^{(j)} } \Bigr)=
 }
  \displaystyle{
 \sum_{a=1}^{N+M} (-1)^{\ps(a)} g_a \Bigl ( e^{\eta (1-2\ps(a))\sum_{j=1}^{n}
 e_{aa}^{(j)} } \Bigr ) =
  }
 \\ \ \\
  \displaystyle{
 =
 \sum_{a=1}^{N+M} (-1)^{\ps(a)} g_a \Bigl ( e^{\eta (1-2\ps(a)){\bf M_a}} \Bigr) =
 }
  \displaystyle{
 \sum_{a \in \mathfrak{B}} g_a e^{ \eta {\bf M}_a} - \sum_{a \in \mathfrak{F}} g_ae^{ -\eta {\bf
 M}_a}\,.\quad \blacksquare
    }
  \end{array}
 \eeq
  Notice that although this expression depends on the choice of $\mathfrak{B}$
  and $\mathfrak{F}$ the eigenvalue of the Ruijsenaars-Schneider Hamiltonian
  is independent of it:
 \beq\label{q440}
   \begin{array}{c}
   \displaystyle{
  \sum_{i=1}^n \left ( \prod_{j\neq i}^n \frac{\sinh (x_i-x_j +\eta )}{\sinh (x_i-x_j)}\right )
e^{\eta \hbar \p_{x_i}}\Psi \,\, =\,\,
\sum_{i=1}^n  \prod_{j\neq i}^n \frac{\sinh (x_i-x_j +\eta )}{\sinh (x_i-x_j)}\,
\Bigl <\Omega_q \Bigr |   {\bf K}_{i}^{(0 )}\Bigl | \Phi \Bigr >
  }
   \\ \ \\
  \displaystyle{
= \sum_{i=1}^n \Bigl <\Omega_q \Bigr |   {\bf H}_{i}\Bigl | \Phi
\Bigr > \,\, = \,\, \Bigl <\Omega_q \Bigr | \frac{{\bf T}( \infty )
- {\bf T}( -\infty )}{2 \sinh \eta}   \Bigl | \Phi \Bigr >
  }
   \\ \ \\
  \displaystyle{
 =
\Bigl <\Omega _q\Bigr |\sum_{a \in \mathfrak{B}} g_a \frac{\sinh (\eta  {\bf M}_a)}{\sinh \eta} +
\sum_{a \in \mathfrak{F}} g_a \frac{\sinh (\eta  {\bf M}_a)}{\sinh \eta}
\Bigl | \Phi \Bigr >
  }
   \\ \ \\
  \displaystyle{
  =
\sum_{a=1}^{N+M} g_a \Bigl <\Omega _q\Bigr | \frac{\sinh (\eta  {\bf
M}_a)}{\sinh \eta} \Bigl | \Phi \Bigr >\,\, =\,\, \left
(\sum_{a=1}^{N+M} g_a \frac{\sinh (\eta M_a )}{\sinh \eta}\right
)\Psi\,.
   }
  \end{array}
 \eeq
  Therefore,
 \beq\label{q492}
   \begin{array}{c}
   \displaystyle{
\Psi = \Bigl<\Omega_q | \Phi\Bigr>
   }
  \end{array}
 \eeq
is indeed an eigenfunction of the Ruijsenaars-Schneider Hamiltonian
with the eigenvalue
 \beq\label{q483}
   \begin{array}{c}
   \displaystyle{
E = \sum_{a=1}^{N+M} g_a \frac{\sinh (\eta M_a )}{\sinh \eta}\,.
   }
  \end{array}
 \eeq

\subsubsection*{Construction of $q$-antisymmetric vectors}

In order to extend the correspondence to the case of the Hamiltonian
with the opposite sign of $\eta$ we should start with a different
$R$-matrix:
 \beq\label{q441}
   \begin{array}{c}
   \displaystyle{
{\bf R}(x)= \frac{1}{2 \sinh (x \! - \! \eta )} \! \sum_{a=1}^{N+M} (e^{x + \eta}
q^{-2\ps (a)} - e^{-x-\eta} q^{2\ps (a)}) e_{aa}\otimes
e_{aa}+\frac{\sinh x}{\sinh (x - \eta )} \sum_{a\neq b}^{N+M}
e_{aa}\otimes e_{bb}
 }
 \\ \ \\
  \displaystyle{
+ \frac{\sinh \eta}{\sinh (x - \eta )}\sum_{a<b}^{N+M} \Bigl ( e^x
(-1)^{\ps(b)} e_{ab}\otimes e_{ba}+e^{-x} (-1)^{\ps(a)}
e_{ba}\otimes e_{ab}\Bigr )\,.
   }
  \end{array}
 \eeq
It is an analog of (\ref{r113}) in the rational case. Expression
(\ref{q441}) can be rewritten in the form
 \beq\label{q442}
   \begin{array}{c}
   \displaystyle{
{\bf R}_{12}(x)\! = -{\bf P}_{12} + \frac{\sinh x}{\sinh (x\! -\!
\eta )}\Bigl ({\bf I}+{\bf P}^q_{12} \Bigr ) + {\bf G}_{12}^-\,,
   }
  \end{array}
 \eeq
where
   \beq\label{q444}
   \begin{array}{c}
   \displaystyle{
{\bf G}_{12}^-    = \sum_{a=1}^{N+M} \Bigl(
\frac{\sinh(x+\eta-2\eta \ps(a))}{\sinh(x-\eta)} + (-1)^{\ps(a)}
-\frac{\sinh(x)}{\sinh(x-\eta)}((-1)^{\ps(a)}+1)\Bigr )e_{aa}\otimes
e_{aa}
  }
   \\ \ \\
  \displaystyle{
 =2\sum_{a \in \mathfrak{B} } \frac{ (\cosh \eta - 1)
\sinh(x)}{\sinh(x-\eta)} \, e_{aa}\otimes e_{aa} = \sum_{a=1}^{N+M}
{\bf G}_a^- \, e_{aa}\otimes e_{aa}\,.
    }
  \end{array}
 \eeq
Similarly to the case of symmetric vector (and also similarly to
(\ref{OMG2})) it is easy to see that the vector
$\Bigl <\Omega_q\Bigr |$ with the property
 \begin{equation}
\Bigl <\Omega_q\Bigr |\,{\bf P}^q_{i,i-1} = - \Bigl <\Omega_q\Bigr |
 \end{equation}
can not contain two or more identical bosonic vectors. On the other hand,  ${\bf G}_{12}^-$
acts by zero on the pair of identical fermions. Thus
 \begin{equation}
\Big <\Omega_q \Big |\,{\bf R}_{i,i-1} = - \Big <\Omega_q \Big |\,{\bf
P}_{i,i-1}\,.
 \end{equation}
Repeating the steps from the previous paragraphs we obtain the
following expressions for the asymptotics of the $R$-matrix  at
infinity:
   \beq\label{q445}
   \begin{array}{c}
      \displaystyle{
 \widetilde{{\bf R}}(\infty) = {\bf I} + (q-q^{-1})\sum_{a>b}^{N+M} (-1)^{\ps(b)}
 e_{ab}\otimes e_{ba} +\sum_{a=1}^{N+M}\Bigl (q^{1-2\ps(a)}-1 \Bigr) e_{aa}\otimes
 e_{aa}\,,
  }
   \\ \ \\
  \displaystyle{
 \widetilde{{\bf R}}(-\infty) = {\bf I} + (q^{-1}-q)\sum_{a<b}^{N+M} (-1)^{\ps(b)}
  e_{ab}\otimes e_{ba} +\sum_{a=1}^{N+M}\Bigl (q^{-1+2\ps(a)}-1 \Bigr) e_{aa}\otimes
  e_{aa}\,,
    }
  \end{array}
 \eeq
 where
  \begin{equation}
 \widetilde{{\bf R}}(x) = \frac{\sinh(x-\eta)}{\sinh x} \, {\bf
 R}(x)\,.
 \end{equation}
 It is easy to see that these asymptotics differ from the corresponding asymptotics
  in the $q$-symmetric case by non-diagonal part only,
 but the latter does not contribute to the trace in the transfer matrix.
 Therefore, the Hamiltonian with the opposite sign of $\eta$ has the same eigenvalue:
  \beq
  \sum_{i=1}^n \left ( \prod_{j\neq i}^n \frac{\sinh (x_i-x_j -\eta )}{\sinh (x_i-x_j)}\right )
e^{\eta \hbar \p_{x_i}}\Psi \,\, =\, \left (\sum_{a=1}^{N+M} g_a
\frac{\sinh (\eta M_a )}{\sinh \eta}\right )\Psi\,.
 \eeq

\subsubsection*{Symmetry between  $q$-(anti)symmetric vectors}

In this paragraph we will show that the usage of $q$-antisymmetric
vectors do not actually lead to any new wave functions of the
Ruijsenaars-Schneider system. For this paragraph let us introduce more refined
notations:
 \begin{equation}\label{tildeR}
 \begin{array}{c}
\displaystyle{\widetilde{{\bf R}}^\ps (x|\eta)=  \frac{1}{2 \sinh x}\sum_{a=1}^{N+M} \left(e^{x+\eta}
q^{-2\ps(a)} - e^{-x-\eta} q^{2\ps(a)}\right) e_{aa}\otimes e_{aa}
+ \sum_{a\neq b}^{N+M}
e_{aa}\otimes e_{bb}}
\\ \\
\displaystyle{+ \frac{\sinh \eta}{\sinh x}\sum_{a<b}^{N+M} \Bigl ( e^x
(-1)^{\ps(b)} e_{ab}\otimes e_{ba}+e^{-x} (-1)^{\ps(a)}
e_{ba}\otimes e_{ab}\Bigr )}
\end{array}
 \end{equation}
 and
 \begin{equation}
{\bf R}^{\ps}_{\pm}(x|\eta) = \frac{\sinh x}{ \sinh(x \pm
\eta)}\, \widetilde{{\bf R}}^{\ps}(x|\eta)\,,
 \end{equation}
 where the index $\ps$ stands for a fixed choice of grading.

 Let us introduce the operator $Q$ of the grading change:
 \begin{equation}
 \ps(Q e_a) = \ps(e_a)+1 \;\; (\mbox{mod $2$}).
 \end{equation}
This operator simply changes all basis bosonic
vectors $e_a$ to fermionic ones and vice versa.
 It is easy to see from this definition that the $R$-matrix has a
symmetry
 \begin{equation}\label{MNsym11}
Q \widetilde{{\bf R}}^{\ps}(x|\eta)Q^{-1} =  \widetilde{{\bf
R}}^{\ps+1}(x|-\eta)\,,
 \end{equation}
where the index
$\ps +1$ means simultaneous shift of all grading parameters by $1$ modulo $2$ in (\ref{tildeR}).
 Therefore,
 \begin{equation}
Q{\bf R}^{\ps}_{-}(x|\eta)Q^{-1} =  {\bf R}^{\ps+1}_{+}(x|-\eta)\,.
 \end{equation}
 For the special vectors (on which we
project the solutions) we  also reserve the following notation:
\beq
\Bigl <\Omega_{q+}^{\ps}\Bigr |\,{\bf P}^{q,\ps}_{i,i-1} =  \Bigl <\Omega_{q+}^{\ps}\Bigr | ,
\quad
\Bigl <\Omega_{q-}^{\ps}\Bigr |\,{\bf P}^{q,\ps}_{i,i-1} = - \Bigl <\Omega_{q-}^{\ps}\Bigr |.
\eeq
 By changing all bosons to fermions in these equations and vice versa, and taking into account that
 \begin{equation}
Q{\bf P}^{q,\ps}_{i,i-1}Q^{-1} = - {\bf P}^{q,\ps+1}_{i,i-1},
 \end{equation}
 we get
 \begin{equation}\label{MNsym21}
\Bigl <\Omega_{q+}^{\ps}\Bigr |Q = \Bigl <\Omega_{q-}^{\ps+1}\Bigr |\,.
 \end{equation}

 As a first step towards the explanation of the origin of the wavefunctions for Hamiltonians with signs of $\eta$ and $\hbar$ changed  we will prove the following
  \begin{predl}
 For any solution $\Bigl |\Phi^{\ps}_{-}(x|\eta,\hbar) \Bigr >$ of the qKZ
 equations  with the R-matrix ${\bf R}^{\ps}_{-}(x|\eta)$ suitable for
 projecting on the $q$-antisymmetric vector $\Bigl <\Omega_{q-}^{\ps}\Bigr |$,
 we can construct the solution $\Bigl |\Phi^{\ps+1}_{+}(x|\eta,\hbar) \Bigr >$ of the qKZ
 equations,  with the R-matrix ${\bf R}^{\ps+1}_{+}(x|\eta)$
 suitable for projecting on the $q$-symmetric vector $\Bigl <\Omega_{q+}^{\ps+1}\Bigr |$.
 \end{predl}
 \noindent\underline{\em{Proof:}}\
 Consider the qKZ-equations:
 \begin{eqnarray*}
e^{\eta \hbar \p_{x_i}} \Bigl | \Phi^\ps_{-}(x|\eta,\hbar) \Bigr >={\bf
R}^{\ps}_{-,i\, i\! -\! 1}(x_i \! -\! x_{i-1}\! +\! \eta \hbar |
\eta)\ldots {\bf R}^{\ps}_{-,i 1}(x_i \! -\! x_{1}\! +\! \eta \hbar
| \eta) {\bf g}^{(i)}  \\
\times {\bf R}^{\ps}_{-,i n}(x_i \!
-\! x_{n} | \eta)\ldots {\bf R}^{\ps}_{-,i\, i\! +\! 1}(x_i \! -\!
x_{i+1} | \eta) \Bigl | \Phi^\ps_{-}(x|\eta,\hbar) \Bigr >, \qquad i=1,
\ldots , n\,.
  \end{eqnarray*}
 Changing signs of $\eta$ and $\hbar$ yields
 \begin{eqnarray*}
e^{\eta \hbar \p_{x_i}} \Bigl | \Phi^\ps_{-}(x|-\eta,-\hbar) \Bigr
>={\bf R}^{\ps}_{-,i\, i\! -\! 1}(x_i \! -\! x_{i-1}\! +\! \eta
\hbar | -\eta)\ldots {\bf R}^{\ps}_{-,i 1}(x_i \! -\! x_{1}\! +\!
\eta \hbar | -\eta) {\bf g}^{(i)}  \\
\times {\bf
R}^{\ps}_{-,i n}(x_i \! -\! x_{n} | -\eta)\ldots {\bf
R}^{\ps}_{-,i\, i\! +\! 1}(x_i \! -\! x_{i+1} | -\eta) \Bigl |
\Phi^\ps_{-}(x|-\eta,-\hbar) \Bigr >, \qquad i=1, \ldots , n\,.
  \end{eqnarray*}
  Using the symmetry (\ref{MNsym11})  this could be rewritten in the form:
  \begin{eqnarray*}
e^{\eta \hbar \p_{x_i}} Q \, \Bigl | \Phi^{\ps}_{-}(x|-\eta,-\hbar) \Bigr
>={\bf R}^{\ps+1}_{+,i\, i\! -\! 1}(x_i \! -\! x_{i-1}\! +\! \eta
\hbar | \eta)\ldots {\bf R}^{\ps+1}_{+,i 1}(x_i \! -\! x_{1}\! +\!
\eta \hbar | \eta) {\bf g}^{(i)}  \\ \times {\bf
R}^{\ps+1}_{+,i n}(x_i \! -\! x_{n} | \eta)\ldots {\bf
R}^{\ps+1}_{+,i\, i\! +\! 1}(x_i \! -\! x_{i+1} | \eta) \, Q\, \Bigl |
\Phi^{\ps}_{-}(x|-\eta,-\hbar) \Bigr >, \qquad i=1, \ldots , n\,.
  \end{eqnarray*}
 It can be seen from here that the desired solution
 $\Bigl |\Phi^{\ps+1}_{+}(x|\eta,\hbar) \Bigr >$ is the following:
  \begin{equation} \label{St}
\Bigl | \Phi^{\ps+1}_{+}(x|\eta,\hbar) \Bigr >   = Q\,  \Bigl |
\Phi^\ps_{-}(x|-\eta,-\hbar) \Bigr >.
  \end{equation}
  $\blacksquare$

Consider the space of all wavefunctions $\Psi_{-}(x|\eta,\hbar)$ of the Ruijsenaars Hamiltonian with signs of  $\eta$ and $\hbar$ changed:
    \beq
  \sum_{i=1}^n \left ( \prod_{j\neq i}^n \frac{\sinh (x_i-x_j -\eta )}{\sinh (x_i-x_j)}\right )
e^{\eta \hbar \p_{x_i}}\Psi_{-}(x|\eta,\hbar) \,\, =\, \left (\sum_{a=1}^{N+M} g_a
\frac{\sinh (\eta M_a )}{\sinh \eta}\right )\Psi_{-}(x|\eta,\hbar)\, ,
 \eeq
 which could be obtained with our construction, i.e. they have the form
 \begin{equation} \label{Psi-}
 \Psi_{-}(x|\eta,\hbar) = \Bigl<\Omega_{q-}^{\ps}\Bigr| \Phi^{\ps}_{-}(x|\eta,\hbar) \Bigr>.
\end{equation}
For any such $\Psi_{-}(x|\eta,\hbar)$ the function  $\Psi_{+}(x|\eta,\hbar) = \Psi_{-}(x|-\eta,-\hbar)$ is automatically satisfies the equation
\beq
  \sum_{i=1}^n \left ( \prod_{j\neq i}^n \frac{\sinh (x_i-x_j +\eta )}{\sinh (x_i-x_j)}\right )
e^{\eta \hbar \p_{x_i}}\Psi_{+}(x|\eta,\hbar) \,\, =\, \left
(\sum_{a=1}^{N+M} g_a \frac{\sinh (\eta M_a )}{\sinh \eta}\right
)\Psi_{+}(x|\eta,\hbar)\,.
 \eeq

Now we are ready to prove the main statement of this section.
 \begin{predl}
For any wavefunction of the form (\ref{Psi-}) the corresponding
$\Psi_{+}(x|\eta,\hbar) = \Psi_{-}(x|-\eta,-\hbar)$ can be also obtained from our
construction, i.e., it has the form
 \begin{equation} \label{Psi+}
 \Psi_{+}(x|\eta,\hbar) = \Bigl <\Omega_{q}^{\ps+1}\Bigr | \Phi^{\ps+1}_{+}(x|\eta,\hbar) \Bigr >.
\end{equation}
  \end{predl}
The proof follows from the previous proposition with
$\Bigl |\Phi^{\ps+1}_{+}(x|\eta,\hbar) \Bigr >$ defined as in (\ref{St}) and the remark (\ref{MNsym21}).

  This proposition actually means that for any wavefunction constructed with the help of
  the $q$-antisymmetric vector the existence of the corresponding solution of the
  qKZ equation
  is a simple consequence of the existence of such solution for the wavefunction with
  signs of $\eta$ and $\hbar$ changed, constructed with the help of the $q$-symmetric vector.

\section{Appendix}
\def\theequation{A.\arabic{equation}}
\setcounter{equation}{0}

Here we give a short summary of notations and definitions related to
the Lie superalgebra $gl(N|M)$.

Let $\mathfrak{B}$ be any one of the subsets of $\{1,2,\dots, N+M\}$
with $ \mathrm{Card} (\mathfrak{B})=N$, and $\mathfrak{F}$ be the
complement set $\mathfrak{F}=\{1,2,\dots, N+M\} \setminus
\mathfrak{B}$. The vector space ${\mathbb C}^{N|M}$ is endowed with
the $\mathbb{Z}_2$-grading.  The grading parameter is defined as
 \beq\label{a0}
 \ps (a)=
 \left\{
 \begin{array}{l}
 0\,,\quad a\in \mathfrak{B} \quad (\hbox{bosons})\,,
 \\
 1\,,\quad a\in \mathfrak{F}\ \quad (\hbox{fermions})\,.
 \end{array}
 \right.
 \eeq
The Lie superalgebra $gl(N|M)$ is defined by the following
relations for the generators $\e_{ab}$:
 \beq\label{a1}
\e_{ab}\e_{cd} -(-1)^{\ps(\e_{ab})\ps(\e_{cd})}\e_{cd}
\e_{ab}=\delta_{bc}\e_{ad}-(-1)^{\ps(\e_{ab})\ps(\e_{cd})}
\delta_{ad}\e_{cb}\,,
 \eeq
 where
 \beq\label{a2}
\ps(\e_{ab})=\ps(a)+\ps(b) \ \hbox{mod} \ 2\,.
 \eeq
 In the fundamental representation the set of generators $\{\e_{ab}\}$
 forms the
 standard basis in matrices ${\rm End}({\mathbb C}^{N|M})$: $(e_{ab})_{ij}=\delta_{ia}\delta_{jb}$, so that
 for the orthonormal basis vectors
 $e_a$, $a=1,...,N+M$ in $\mathbb{C}^{N|M}$ (i.e. $(e_a)_k=\delta_{ak}$) we have
\beq\label{a333}
e_{ab}\, e_c=\delta_{bc}\, e_a\,.
 \eeq

For any homogeneous (with a definite grading) operators $\{
\mathbf{A}_{i}\in {\rm End}({\mathbb C}^{N|M}) \}_{i=1}^{4}$ and
homogeneous vectors $\mathbf{x}\,,\mathbf{y}\in \mathbb{C}^{N|M}$ we have:
  \beq\label{a40}
(\mathbf{A}_{1} \otimes \mathbf{A}_{2}) (\mathbf{x}\otimes
\mathbf{y})= (-1)^{\ps(\mathbf{A}_{2})\ps(\mathbf{x})}
(\mathbf{A}_{1}\mathbf{x} \otimes \mathbf{A}_{2}\mathbf{y})
 \eeq
and
  \beq\label{a4}
(\mathbf{A}_{1} \otimes \mathbf{A}_{2}) (\mathbf{A}_{3} \otimes
\mathbf{A}_{4})= (-1)^{\ps(\mathbf{A}_{2})\ps(\mathbf{A}_{3})}
(\mathbf{A}_{1}\mathbf{A}_{3} \otimes
\mathbf{A}_{2}\mathbf{A}_{4})\,.
 \eeq
The graded permutation operator $\mathbf{P}_{12}\in \mathrm{End}
\bigl (\mathbb{C}^{N|M}\otimes \mathbb{C}^{N|M}\bigr )$ is of the
form:
  \beq\label{a5}
\mathbf{P}_{12}= \sum_{a,b=1}^{M+N} (-1)^{\ps(b)} e_{ab}\otimes
e_{ba}.
 \eeq
Due to (\ref{a40}) it permutes any pair of homogeneous vectors
$\mathbf{x}$ and $\mathbf{y}$ according to the rule
  \beq\label{a6}
\mathbf{P}_{12}\, \mathbf{x}\otimes \mathbf{y}= (-1)^{\ps
(\mathbf{x}) \ps (\mathbf{y})} \mathbf{y}\otimes \mathbf{x}\,.
 \eeq
In particular,
  \beq\label{a60}
{\bf P}_{12}\, e_{a}\otimes e_{a} = (-1)^{\ps (a)}\, e_{a}\otimes
e_{a}\,.
 \eeq
The supertrace and the superdeterminant of ${\cal M}\in {\rm
End}({\mathbb C}^{N|M})$ are given by
  \beq\label{a7}
\mathrm{str}\, {\cal M}=\sum\limits_{a=1}^{N+M} (-1)^{\ps (a)} {\cal
M}_{aa}
 \eeq
 and $\mathrm{sdet}\, {\cal M}=\exp (\mathrm{str} \log {\cal M})$.
For an operator ${\cal M}^{(i)}$ acting as ${\cal M}$ on the $i$-th
component of $({\mathbb C}^{N|M})^{\otimes n}$ we have
  \beq\label{a9}
{\bf P}_{ij} \, {\cal M}^{(j)} = {\cal M}^{(i)} \, {\bf P}_{ij}\,,
 \eeq
  \beq\label{a10}
\mbox{str}_0 ({\bf P}_{0i}\, {\cal M}^{(0)}) = {\cal M}^{(i)}\,.
 \eeq

\section*{Acknowledgments}

The work of A. Grekov was supported in part by RFBR grant
18-02-01081. The work of A. Zabrodin and A. Zotov has been funded by
the Russian Academic Excellence Project `5-100'. The work of A.
Zotov was also supported in part by RFBR grant 18-01-00273.

\begin{small}
 
 \end{small}
 \end{document}